\documentclass[a4paper]{jpconf}
\usepackage{graphicx}
\begin{document}
\title{Correlations and fluctuations studied with ALICE}

\author{Michael Weber for the ALICE collaboration}
\address{University of Houston}

\ead{m.weber@cern.ch}

\begin{abstract}

The measurement of particle correlations and event--by--event fluctuations of physical observables allows to study a large variety of properties of the matter produced in ultra relativistic heavy--ion collisions. We will present results for two--particle correlations, mean transverse momentum fluctuations, and net charge fluctuations in Pb--Pb collisions at $\sqrt{s_{\mathrm {NN}}}= 2.76$ TeV. 
 
\end{abstract}

\section{Motivation}

In ultra relativistic heavy--ion collisions the created nuclear matter reaches high temperatures, with a sufficient energy density to produce a deconfined state of quarks and gluons. The study of this so-called quark gluon plasma (QGP) as predicted by quantum chromodynamics (QCD) \cite{Ref:QCD} is the aim of the ALICE experiment at the Large Hadron Collider (LHC) at CERN \cite{Ref:AlicePPR,Ref:AliceJinst}. At RHIC (Relativistic Heavy--Ion Collider) energies evidence for the existence of this new state of matter has already been published \cite{Ref:RHICQGP}. Entering a new energy regime with the LHC heavy--ion program, the existence of the QGP has further been established by first experimental results \cite{Ref:AliceFlow,Ref:LHCHighPt}.
The questions that now naturally emerge in this context can be summarized as: 
\begin{itemize}
\item What are the properties of the QGP, e.g. $\eta/s$, medium transport coefficients? 
\item What is the evolution from QGP formation to final state hadrons which are measured in the experiments, e.g. thermalization, phase transition, hadronization?
\end{itemize}
The study of correlations among the produced particles and the event--by--event fluctuations of quantities such as the net charge and the mean transverse momentum is proposed to provide some of the answers. We will present with the help of two examples the strength of these methods to learn about the properties and evolution of the QGP. The analyses are performed on data taken in Pb--Pb collisions at $\sqrt{s_{\mathrm {NN}}}= 2.76$ TeV making use of the excellent tracking capabilities of the ALICE Time Projection Chamber (TPC). The primary vertex is provided by both the TPC and the Silicon Pixel Detector (SPD) and the centrality of the collisions is determined by scintillation detectors (VZERO-A and VZERO-C) as described in \cite{Ref:ALICECentrality}.

\newpage

\section{Correlations}

\subsection{Particle yield modification in jet--like azimuthal dihadron correlations}

\begin{figure*}[htb]
\centering
\includegraphics[width=1.0\linewidth]{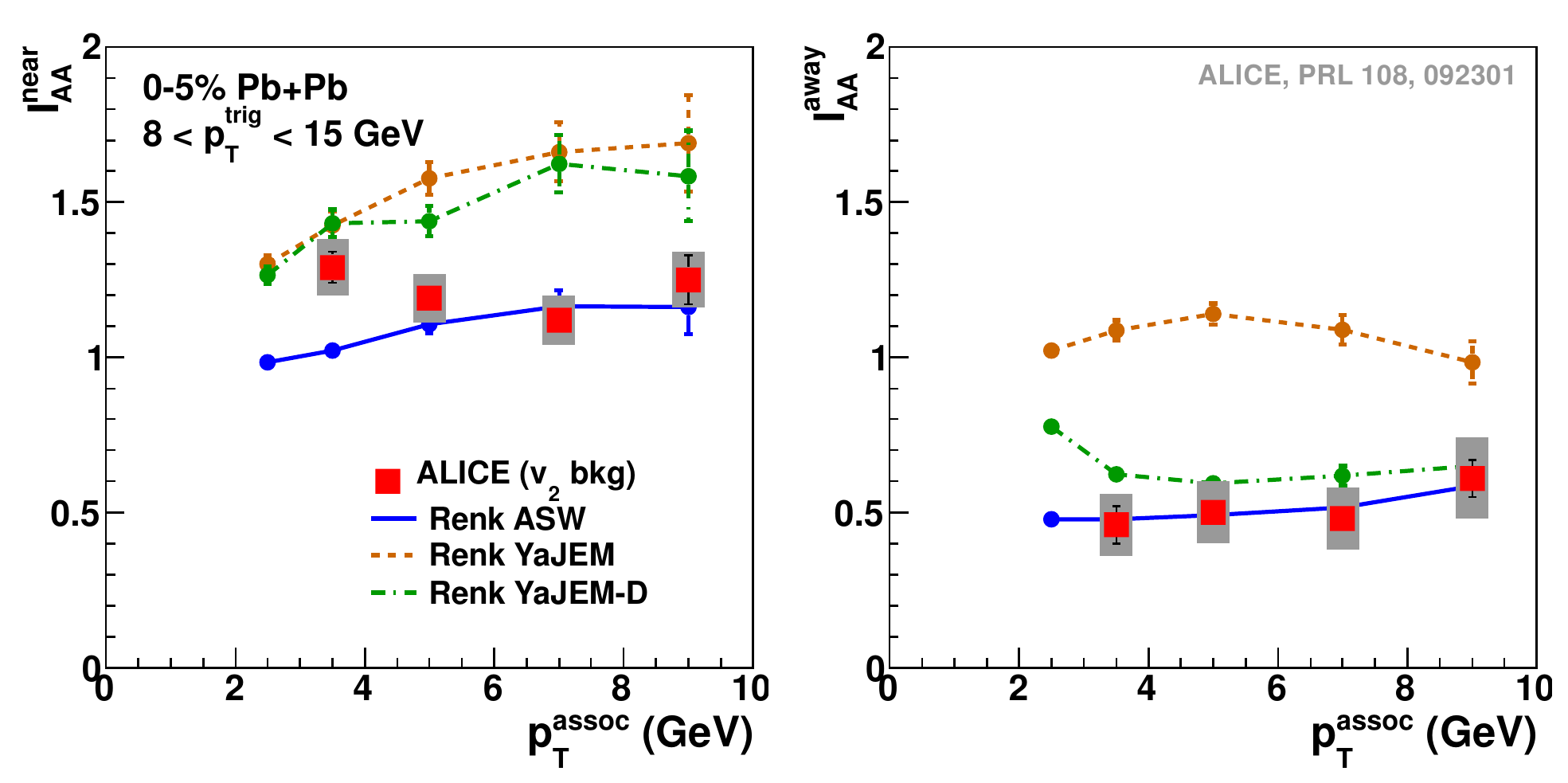}
\caption{Charged hadron $I_{\mathrm {AA}}$ for trigger particle $p_{\rm T} = 8-15$ GeV/$c$ \cite{Ref:ALICEIAA} compared to different shower evolution models \cite{Ref:Renk}.}
\label{fig:IAA:Renk}
\end{figure*}

One way to learn about the properties of the QGP is to study the in--medium energy loss of high transverse momentum partons, which are produced in initial hard scattering processes \cite{Ref:JetQuenching} and are measurable in the suppression of high--$p_{\rm T}$ hadrons in the nuclear modification factor $R_{\mathrm {AA}}$ \cite{Ref:LHCHighPt}. Further constraints on the quenching mechanisms can be retrieved from the per trigger yield of associated particles \cite{Ref:ALICEIAA}, which is defined for azimuthal dihadron correlations as
\begin{equation}
C(\Delta\varphi) = \frac{1}{N_{trig}}\frac{dN_{assoc}}{d\Delta\varphi}.
\end{equation}
After background subtraction the yield Y is integrated for momentum intervals, where inter-- and intra--jet correlations are the dominating source of particle correlations ($8$~GeV/$c<p_{\rm T,trig}<15$~GeV/$c$ and $3$~GeV/$c<p_{\rm T,assoc}<p_{\rm T,trig}$) in two regions: the near--side peak ($|\Delta \varphi| < 0.7$~$rad$) and the away--side peak ($|\Delta \varphi-\pi| < 0.7$~$rad$). The ratio of yields in central Pb--Pb to pp collisions $I_{\mathrm {AA}}$ is shown in Fig.~\ref{fig:IAA:Renk} in comparison to different shower evolution models \cite{Ref:Renk}. In the away side a strong suppression ($I_{\mathrm {AA}}\simeq0.6$) of associated particles due to quenching is observed. At the near--side a moderate enhancement ($I_{\mathrm {AA}}\simeq1.2$) is observed, which can be understood as (i) a change of the fragmentation function, (ii) a possible change of the quark/gluon jet ratio in the final state due to the different coupling to the medium, and (iii) a bias on the parton $p_{\rm T}$ spectrum after energy loss due to the trigger particle selection. Only one model (ASW) is able to reproduce both, the near-- and the away--side $I_{\mathrm {AA}}$, showing the strong constraints that can be obtained from this measurement. 

\subsection{Harmonic decomposition}

\begin{figure*}[htb]
\includegraphics[width=0.5\linewidth]{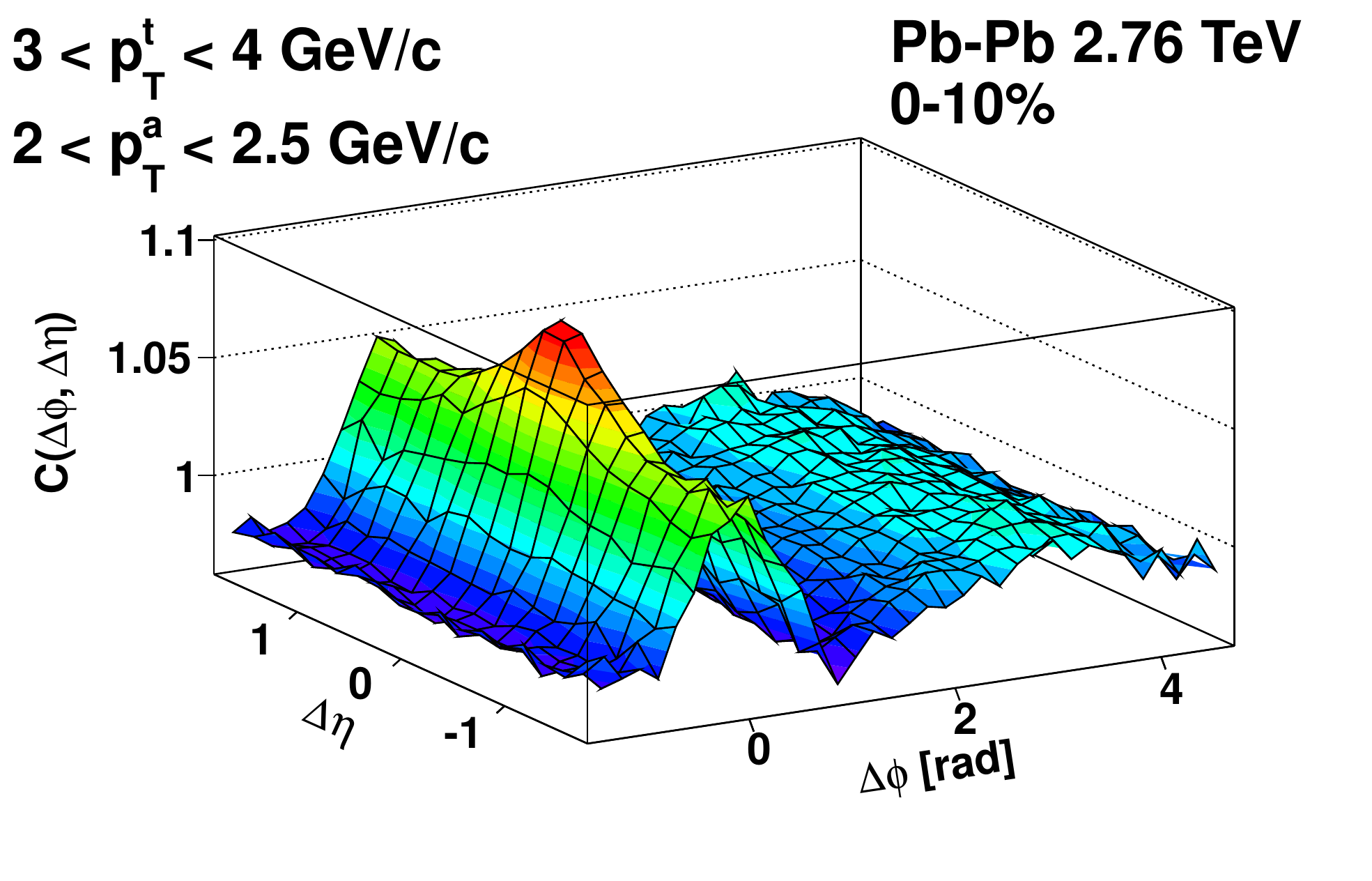}
\includegraphics[width=0.5\linewidth]{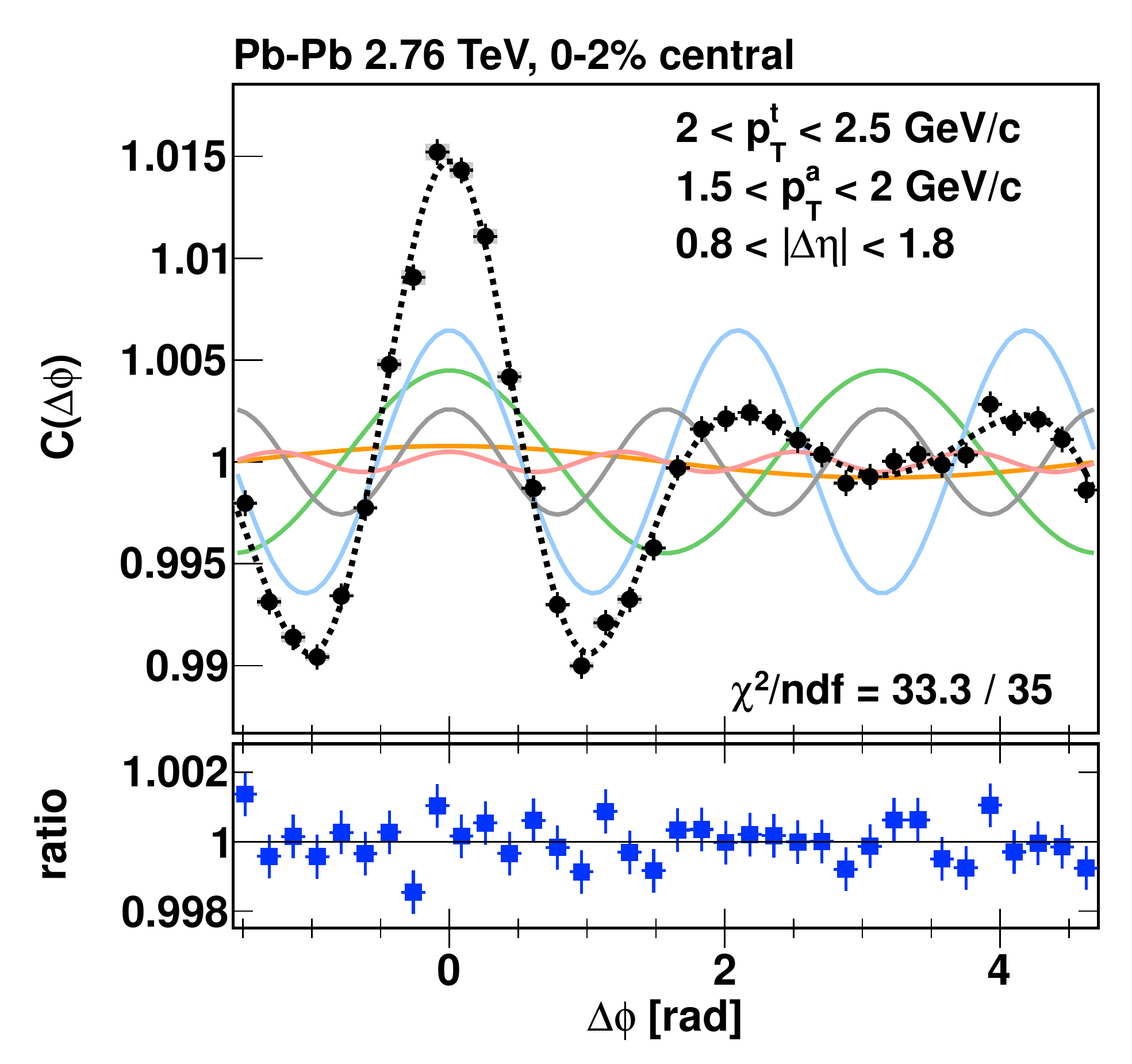}
\caption{Left panel: Example of two--particle correlation functions for central Pb--Pb collisions at low to intermediate transverse momentum. Right panel: $C (\Delta\varphi)$ for particle pairs at $|\Delta\eta| > 0.8$. The Fourier harmonics for $V_{1\Delta}$ to $V_{5\Delta}$ are superimposed in color. Their sum is shown as the dashed curve. The ratio of data to the $n\leq5$ sum is shown in the lower panel. Figure from \cite{Ref:ALICEHarmonics}.}
\label{fig:harmonic:single}
\end{figure*}

\begin{figure*}[htb]
\includegraphics[width=\linewidth]{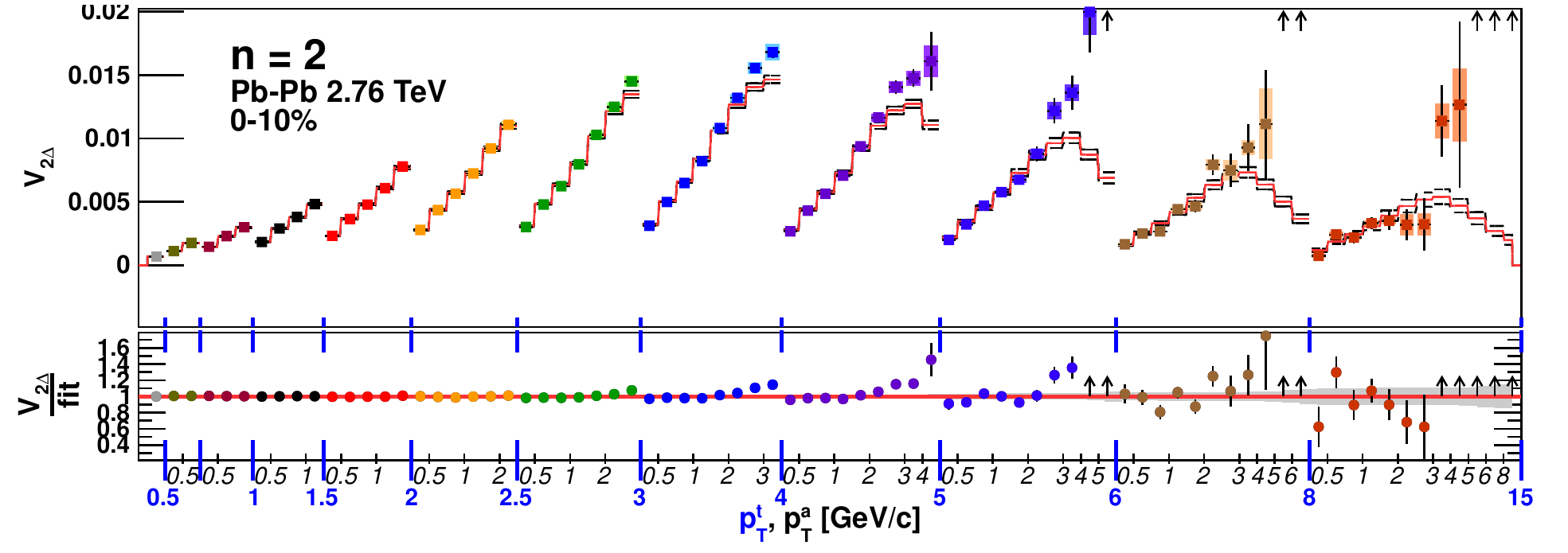}
\caption{Global fit examples in 0--10\% central events for n = 2. The measured $V_{2\Delta}$ coefficients are plotted on an interleaved $p_{\rm T,trig}$, $p_{\rm T,assoc}$ axis in the upper panels, and the global fit function is shown as the red curves. The global fit systematic uncertainty is represented by dashed lines. The lower section shows the ratio of the data to the fit, and the shaded bands represent the systematic uncertainty propagated to the ratio. In all cases, off--scale points are indicated with arrows. Figure from \cite{Ref:ALICEHarmonics}.}
\label{fig:harmonic:globalfit}
\end{figure*}

Moving from the jet--like particle correlations to the soft regime, where collective processes are the dominant source of particle correlations, the question of initial conditions in the heavy--ion collision can be tackled. To study the shape evolution of triggered particle pair distributions in $\Delta\varphi$ quantitatively \cite{Ref:ALICEHarmonics}, the correlation function 
\begin{equation}
C(\Delta\eta,\Delta\varphi) = \frac{N_{mixed}}{N_{same}} \frac{N_{same}(\Delta\eta,\Delta\varphi)}{N_{mixed}(\Delta\eta,\Delta\varphi)}
\end{equation}
was calculated from the distributions of particle pairs formed from the same event $N_{same}(\Delta\eta,\Delta\varphi)$ and from different events $N_{mixed}(\Delta\eta,\Delta\varphi)$ for different trigger and associated particle transverse momentum ranges. The ratio of mixed-event $N_{mixed}$ to same-event $N_{same}$ pair counts is included as a normalization factor. 
In the bulk dominated regime ($p_{\rm T} < 3-4$~GeV/$c$) in addition to the near-- and away--side peak at small pseudorapidity intervals between trigger and associate particle a near-side ridge and a peaked structure are observed for $0.8<\Delta\eta<1.8$ (see left panel of Fig.~\ref{fig:harmonic:single}). The decomposition in Fourier components, both even and odd:
\begin{equation}
V_{n\Delta} = \langle \cos n\Delta\varphi \rangle = \frac{\int{d\Delta\varphi C (\Delta\varphi) cos n\Delta\varphi  }}{\int{d\Delta\varphi C (\Delta\varphi) }}
\end{equation}
are shown in the right panel of Fig.~\ref{fig:harmonic:single}. It is sufficient to use components up to $n=5$ to describe the correlation distributions. Furthermore, it is found that for $p_{\rm T,assoc}<4$~GeV/$c$ the two--particle Fourier coefficients $V_{n\Delta}$ are consistent with the single-particle harmonic coefficients obtained from a global fit, except for $n=1$. In Fig.~\ref{fig:harmonic:globalfit} this is shown for $n=2$ as an example.  In addition, the $\mathsf{v}_n$ values obtained from the global fit agree with two--particle cumulant measurements for $n=2-5$ \cite{Ref:ALICEHigherHarmonics}. This decomposition can be understood as the collective response of produced particles to anisotropic initial conditions. However, this decomposition is not necessarily unique to hydrodynamic flow. Factorization may be observed whenever both particles are correlated to one another through their correlation to a common symmetry plane having large longitudinal extent.

\section{Fluctuations}

\subsection{Mean--$p_T$ fluctuations}

It was proposed in \cite{Ref:Stephanov,Ref:Gavin} to use event--by--event fluctuations of the mean transverse momentum in heavy--ion collisions as an indicator of the critical behaviour of the system in the vicinity of a phase transition. Furthermore the influence of early--time dynamics like geometric fluctuations of the colliding system can leave their imprint on these fluctuations. As a reference pp collisions are used, where the influence of known physics, like resonance decays, HBT effects, (mini--) jets can be studied. The two--particle correlator $C_m$ for mean--$p_T$ fluctuations is defined by the covariances of all pairs of particles i and j in the same event with respect to the inclusive mean transverse momentum $\langle p_{\rm T} \rangle_m$ in a multiplicity class $m$:
\begin{equation}
\sigma_{dyn}^{2} \equiv C_{m} = \langle \Delta p_{\rm T,i} \Delta p_{\rm T,j} \rangle = \frac{1}{\sum_{k=1}^{N_{ev}}{N_{k}^{pairs}}} \cdot \sum_{k=1}^{N_{events}} \sum_{i=1}^{N_{k}} \sum_{j=i+1}^{N_{k}} (p_{\rm T,i}-\langle p_{\rm T} \rangle_m)(p_{\rm T,j}-\langle p_{\rm T} \rangle_m)
\end{equation}
where $N_k$ is the number of particles and $N_{k}^{pairs} = 0.5\cdot N_k(N_k-1)$ is the number of particle pairs in event k. It is a measure of the dynamical fluctuations and $C_m$ vanishes in the presence of only statistical fluctuations.
\begin{figure*}[htb]
\centering
\includegraphics[width=0.7\linewidth]{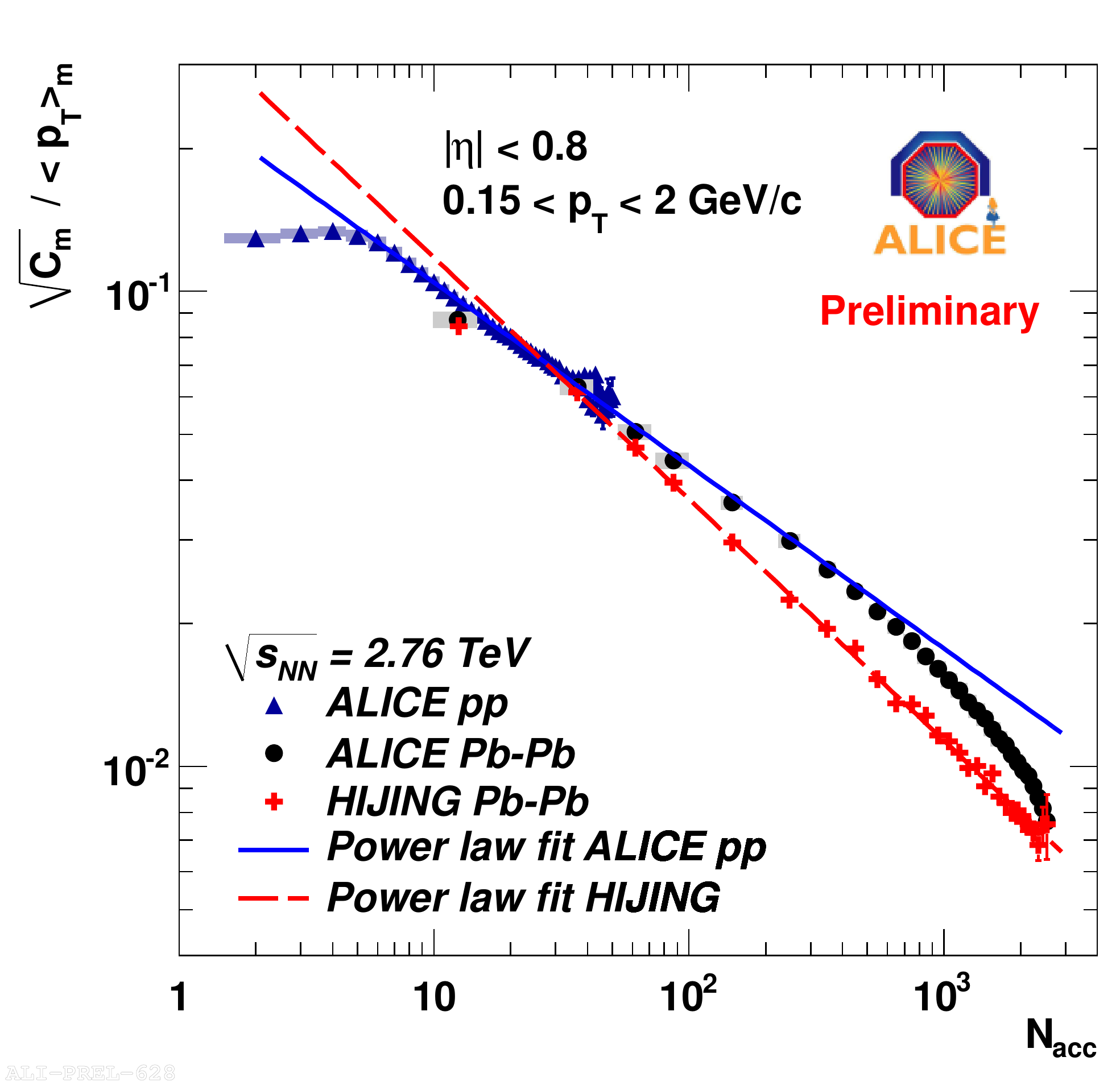}
\caption{Comparison of relative mean transverse momentum fluctuations $\sqrt{C_m}/\langle p_{\rm T} \rangle_m$ in pp and Pb--Pb collisions together with a HIJING simulation and power law fits both for pp data and HIJING simulation \cite{Ref:HIJING}. }
\label{fig:pTfluc}
\end{figure*}

In Fig.~\ref{fig:pTfluc} the relative fluctuations, normalized to the inclusive mean transverse momentum $\langle p_{\rm T} \rangle_m$, are shown as a function of the uncorrected event multiplicity $N_{acc}$, the number of tracks accepted by the analysis cuts. In both pp and Pb--Pb collisions significant non--statistical fluctuations are observed, diluting with increasing multiplicity. A common scaling in pp and peripheral Pb--Pb can be described by a power law fit $\sqrt{C_m}/\langle p_{\rm T} \rangle_m (N_{acc}) = A \cdot N_{acc}^b$ to the pp data with $b=-0.385\pm 0.003$, whereas for higher multiplicities in Pb--Pb collisions an additional reduction of mean transverse momentum fluctuations is evident. This can be understood in the onset of local thermal equilibrium at higher centralities \cite{Ref:Gavin}. The results from the HIJING event generator \cite{Ref:HIJING}, on the contrary, can be described by a single power law fit parameter, differing from the one extracted from the data, which is in agreement with a scaling by $1/\sqrt{N_{acc}}$.

\subsection{Net charge fluctuations}

\begin{figure*}[htb]
\centering
\includegraphics[width=0.7\linewidth]{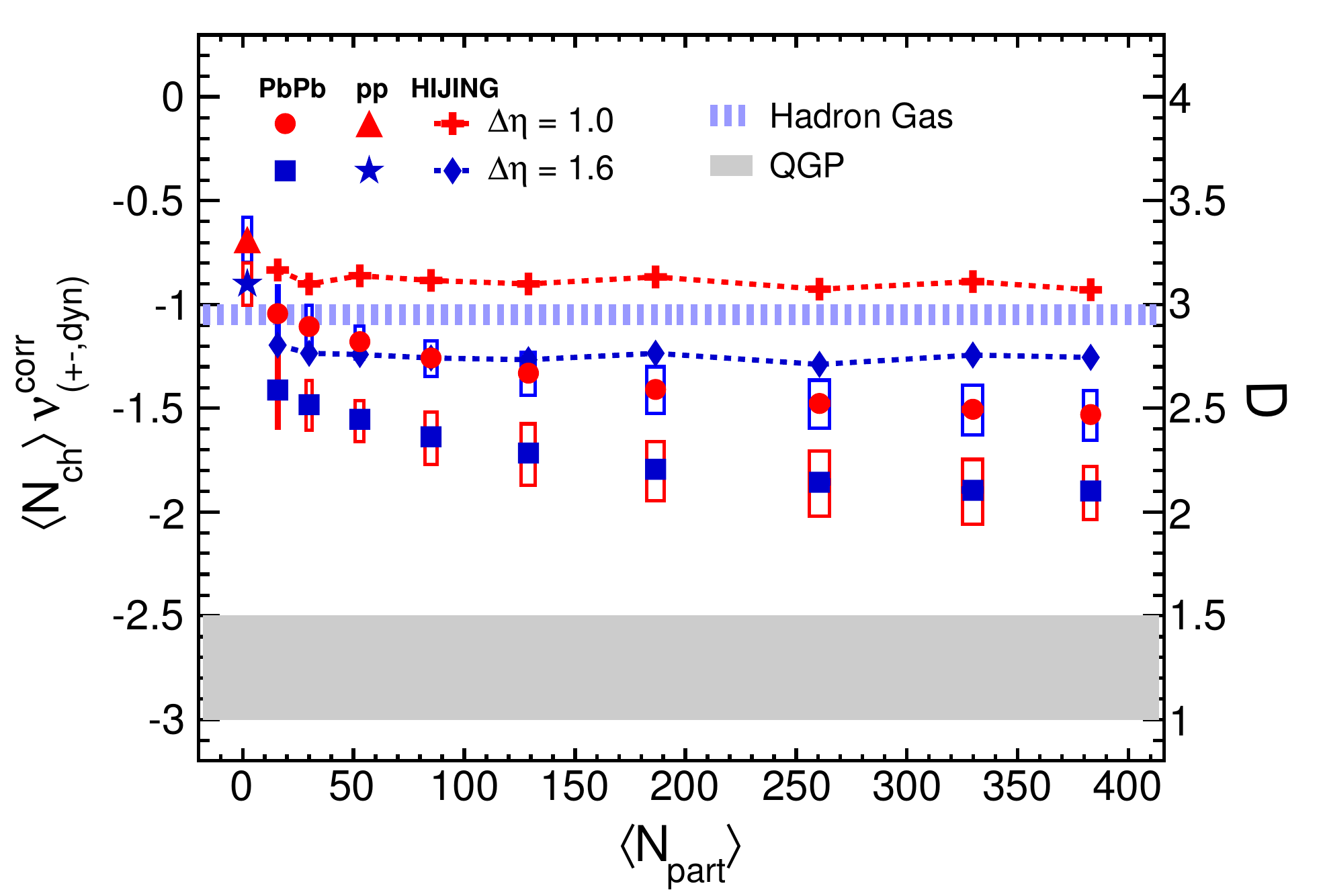}
\caption{$\nu_{+-,dyn}^{corr} \langle N_{ch} \rangle$ (left axis) and $D$ (right axis) as a function of the number of participants for $\Delta\eta=1$ and $\Delta\eta=1.6$ in Pb--Pb collisions at $\sqrt{s_{\mathrm {NN}}}=2.76$~TeV. Also shown are the results from the HIJING event generator for both the $\Delta\eta$ windows and in the shaded bands the expectations for a hadron resonance gas and the QGP \cite{Ref:chargeFluctuations2}. Figure from \cite{Ref:NetChargeFluctuations}.}
\label{fig:chargeFluc:nudyn}
\end{figure*}

Deeper insight on the phase transition to a QGP phase can be gained by determining the relevant charge carriers of the system produced in a heavy--ion collision \cite{Ref:NetChargeFluctuations}. In a hadron (resonance) gas these are mesons and baryons, in the QGP phase quarks are carrying the electric charge. Using the number of positive $N_+$ and negative particles $N_-$ in
\begin{equation}
\nu_{+-,dyn}=\frac{\langle N_+(N_+-1)\rangle}{\langle N_+\rangle^2}+\frac{\langle N_-(N_--1)\rangle}{\langle N_-\rangle^2}-2\frac{\langle N_-N_+\rangle}{\langle N_-\rangle\langle N_+\rangle}
\end{equation}
as a measure for the event-by-event net charge fluctuations, a significant difference between the two phases is expected. The magnitude of the fluctuations can be expressed by the quantity D, defined as \cite{Ref:chargeFluctuations1}
\begin{equation}
D=4\frac{\langle\delta Q^2\rangle}{N_{ch}} \approx \nu_{+-,dyn}\langle N_{ch}\rangle + 4
\end{equation}
with $\langle\delta Q^2\rangle$ the variance of the net charge $Q$ with $Q=N_{+}+N_{-}$ and $N_{ch}=N_{+}+N_{-}$. 
In Fig.~\ref{fig:chargeFluc:nudyn} the values of $\nu_{+-,dyn}^{corr}\langle N_{ch} \rangle$, corrected for global charge conservation, are shown for pp and Pb--Pb collisions as a function of number of participating nucleons $N_{part}$ in a pseudorapidity window $\Delta\eta = 1$ and $\Delta\eta = 1.6$. The shaded bands indicate the predictions for a hadron resonance gas (HG) and the QGP \cite{Ref:chargeFluctuations2}. The data points show a monotonic decreasing centrality dependence and lie clearly below the HG, but above the QGP expectation. On the other hand, the results from  the HIJING event generator for both pseudorapidity windows do not show any dependence on $N_{part}$ and are found in the vicinity of the HG band.  The observed pseudorapidity dependence of $\nu_{+-,dyn}^{corr} \langle N_{ch} \rangle$ may hint to a dilution of the primordial fluctuations during the evolution of the system from hadronization to kinetic freeze--out because of the diffusion of charged hadrons in rapidity \cite{Ref:FlucDiffusion1,Ref:FlucDiffusion2}.

\section{Summary and outlook} 

In summary, it was shown that the ALICE experiment offers a large variety of tools to study the properties and evolution of the QGP via particle correlations and event--by--event fluctuations. Energy--loss models can be constrained, initial conditions, composition, and evolution of the matter produced in ultra relativistic heavy--ion collisions can be probed. To further improve our understanding many more observables or extensions of the presented analyses are needed. Among them, we are studying three--particle correlations, harmonic decomposition on an event--by--event basis, $\Delta p_T \Delta p_T$ correlations, and the charge correlations using the balance function. 



\section*{References}
\begin{thebibliography}{9}

\bibitem{Ref:QCD} {H.~Satz, Rep. Prog. Phys. \textbf{63}, (2000) 1511; \\
S.A.~Bass, M.~Gyulassy, H.~St\"{o}cker, W.~Greiner, J. Phys. \textbf{G25}, (1999) R1; \\
E.V.~Shuryak, Phys. Rep. \textbf{115}, (1984) 151; \\
J.~Cleymans, R.V.~Gavai, E.~Suhonen, Phys. Rep. \textbf{130}, (1986) 217.}


\bibitem{Ref:AlicePPR} {K.~Aamodt \textit{et al.} [ALICE Collaboration], J. Phys. \textbf{G30}, (2004) 1517; \\
K.~Aamodt \textit{et al.} [ALICE Collaboration], J. Phys. \textbf{G32}, (2006) 1295.}

\bibitem{Ref:AliceJinst} {K. Aamodt \textit{et al.} [ALICE Collaboration], JINST \textbf{3}, (2008) S08002.}

\bibitem{Ref:RHICQGP} {
I.~Arsene \textit{et al.} [BRAHMS Collaboration], Nucl. Phys. \textbf{A757}, (2005) 1. \\
K.~Adcox \textit{et al.} [PHENIX Collaboration], Nucl. Phys. \textbf{A757}, (2005) 184. \\
B.~B.~Back \textit{et al.} [PHOBOS Collaboration], Nucl. Phys. \textbf{A757}, (2005) 28. \\
J.~Adams \textit{et al.} [STAR Collaboration], Nucl. Phys. \textbf{A757}, (2005) 102.}

\bibitem{Ref:AliceFlow} {K.~Aamodt \textit{et al.} [ALICE Collaboration], Phys. Rev. Lett. \textbf{105}, (2010) 252302. \\
K.~Aamodt \textit{et al.} [ALICE Collaboration], Phys. Rev. Lett. \textbf{107}, (2011) 032301.}

\bibitem{Ref:LHCHighPt} {
G.~Aad \textit{et al.} [ATLAS Collaboration], Phys. Rev. Lett. \textbf{105}, (2010) 252303.\\
K.~Aamodt \textit{et al.} [ALICE Collaboration], Phys. Lett. \textbf{B696}, (2011) 30.}

\bibitem{Ref:ALICECentrality} {
K.~Aamodt \textit{et al.}, Phys. Rev. Lett. \textbf{106}, (2011) 032301.
}

\bibitem{Ref:JetQuenching}{
	J. D. Bjorken, FERMILAB-PUB-82-059-THY (1982).\\
	M. Gyulassy and M. Plumer, Phys. Lett. B 243, (1990) 432.\\
	X.-N. Wang and M. Gyulassy, Phys. Rev. Lett. 68, (1992) 1480.
	}

\bibitem{Ref:ALICEIAA}{
K.~Aamodt \textit{et al.} [ALICE Collaboration], Phys. Rev. Lett. \textbf{108}, (2012) 092301.
}

\bibitem{Ref:Renk}{
T.~Renk, K.~Eskola, \textit{arXiv:}1106.1740.
}

\bibitem{Ref:ALICEHarmonics}{
K.~Aamodt \textit{et al.} [ALICE Collaboration], Phys. Lett. \textbf{B708}, (2012) 249.
}

\bibitem{Ref:ALICEHigherHarmonics}{
K.~Aamodt \textit{et al.} [ALICE Collaboration],  Phys. Rev. Lett. \textbf{107}, (2011) 032301.
}

\bibitem{Ref:Stephanov}{
    M.A.~Stephanov, K.~Rajagopal and E.V.~Shuryak, Phys.Rev.Lett. \textbf{81}, (1998) 4816-9.\\
    M.A.~Stephanov, K.~Rajagopal and E.V.~Shuryak, Phys.Rev. \textbf{D60}, (1999) 114028.
}

\bibitem{Ref:Gavin}{
    S.~Gavin, Phys. Rev. Lett. \textbf{92}, (2004) 162301.
}

\bibitem{Ref:HIJING}
{M.~Gyulassy and X.~N.~Wang, Comput. Phys. Commun. \textbf{83}, 307 (1994). \\
X.~N.~Wang and M.~Gyulassy, Phys. Rev. \textbf{D44}, 3501 (1991).}

\bibitem{Ref:NetChargeFluctuations} {
B.~Abelev \textit{et al.}, \textit{arXiv:}1207:6068.
}

\bibitem{Ref:chargeFluctuations1}{
S.~Jeon, V.~Koch, Phys. Rev. Lett. 85 (2000), 2076.
}

\bibitem{Ref:chargeFluctuations2}{
S.~Jeon and V.~Koch, In Quark–Gluon–
Plasma 3, Ed. R.C. Hwa and X.N. Wang, 430 (2004);
arXiv:hep-ph/0304012v1.
}

\bibitem{Ref:FlucDiffusion1}{
E. V. Shuryak, M. A. Stephanov, Phys. Rev. C \textbf{63}, (2001) 064903.
}

\bibitem{Ref:FlucDiffusion2}{
M. A. Aziz, S. Gavin, Phys. Rev. C \textbf{70}, (2004) 034905.
}


\end{thebibliography}
\end{document}